\documentclass[letterpaper,twocolumn,10pt]{article}
\usepackage{usenix2024_SOUPS}
\usepackage{balance}

\usepackage{tikz}
\usepackage{amsmath}

\usepackage{filecontents}

\begin{document}

\date{}

\title{\Large \bf Mobile Apps vs. Web Browsers: A User Perception Study with Android Apps and Google Chrome}

\def\plainauthor{AI Decision Making}

\author{
{\rm Harel Berger}\\
Ariel University\\
harelb@ariel.ac.il
} 


\maketitle
\thecopyright



\begin{abstract}
This study examines user perceptions of mobile applications (apps) versus web browsers for accessing online services, with an emphasis on security, privacy, and usability aspects.
Through a combination of an experiment and a survey with Android smartphone users, the research seeks to identify the key concerns and preferences that influence their choice between mobile apps and web browsers. The findings will offer valuable insights for developers to improve the security, privacy and usability of both platforms by addressing user concerns and misconceptions.
\end{abstract}

\section{Introduction}

In today's digital landscape, numerous online services are accessible via smartphones through mobile apps, web browsers, or both.
Although there are mutual threats across platforms, such as attacks on anti viruses~\cite{song2021advanced,berger2021crystal,berger2022problem,xu2014evasion,berger2023breaking} each platform offers distinct advantages and disadvantages on various aspects, including security, privacy and usability~\cite{8835286}. Mobile apps can be a platform of broad malicious activities, as their content includes different types of files~\cite{du2018empirical}. No explicit download of additional files is needed to get hacked besides the app itself. However, mobile operating systems and apps may use advanced security features like biometrics authentication~\cite{zhang2019android}. Web browsers, relying on cookies for session management, can be susceptible to web attacks~\cite{carroll2023human,10.1145/1125451.1125615,sharevski2024exploring,gupta2017cross}.
However, significant attack on the host machine cannot be acheived soley through webservers without additional downloaded files.

Privacy concerns also differ. Mobile apps request access to sensitive data and actions, controlled by user permissions, while web browsers pose risks through tracking and data collection~\cite{8013420}.

Regarding usability, mobile apps integrate better with device features like cameras, GPS, and notifications, and they can work offline, while it is rare to find certain types of websites that would be usable also in offline mode~\cite{4725734}. 
Web browsers, however, do not require installation and updates, offering simpler use. 

This study examines user perceptions of mobile apps versus web browsers, focusing on security, privacy and usability, with Android apps and Google Chrome as case studies. The findings will help developers enhance security, privacy and usability. By comparing these platforms, the study will reveal user misconceptions about the features of these platforms. The findings will show contexts where one platform is preferred, aiding service providers in meeting user needs and creating more user-friendly services.

\section{Research Questions}
\textbf{RQ1:} What security concerns influence users' choice between mobile apps and web browsers for online services?
\\
\textbf{RQ2:} How do users' privacy concerns differ between mobile apps and web browsers?
\\
\textbf{RQ3:} What usability features do users prefer in mobile apps versus web browsers?\\
\textbf{RQ4:} How do users' concerns and preferences differ across various online service categories (e.g., banking, e-shopping, news, social media) between mobile apps and web browsers?

\section{Methodology}

This study will utilize two research methods: an experiment and a survey.

\textbf{Participants:} Approximately 100 students who use Android smartphones will participate in the experiment.

\textbf{Experiment:} The experiment will involve participants performing specific tasks using both mobile apps and web browsers, employing the think-aloud protocol to capture real-time thoughts and reactions. Scenarios will be designed to mimic realistic usage of online services. For example, participants will be asked to complete several tasks using their accounts, to be as authentic as possible: (1) Perform a banking transaction (e.g., transferring money) using a banking app and the web version of the bank’s service; (2) Purchase an item from an online store using both the mobile app and the web browser; (3) Access a news website and read an article using both the mobile app and the web browser.

During these tasks, participants will verbalize their thoughts, describing their actions, any security warnings or alerts encountered, and steps taken to secure their actions (e.g., use of biometrics, password entry). Their actions and difficulties will be recorded for further analysis.

\textbf{Survey:}
Following the experiment, participants will complete a survey to capture their perceptions and concerns regarding the use of mobile apps versus web browsers. Some questions will require participants to respond using a Likert scale, ranging from 1 to 10.\\
\textit{Example Survey Questions:}\\
1. How concerned are you about malware when downloading mobile apps?\\
2. How effective do you find browser security warnings (e.g., untrusted website warnings)?\\
3. How comfortable are you with the permissions requested by mobile apps (e.g., access to contacts, location)?\\
4. Do you feel more in control of your privacy settings in mobile apps or web browsers?\\
5. Which platform do you find easier to navigate: mobile apps or web browsers?\\
6. How important is offline access for you when using an online service?\\
7. Does the available storage on your device influence your decision to download a new app?
\\Additionally, demographic information such as age and language proficiency will be collected from participants.

\textbf{Data Analysis:} The data from the experiment will be analyzed to identify patterns and issues related to security, privacy, and usability. The analysis will be categorized based on services, such as e-shopping, news, direct messages, etc.

Qualitative data from the think-aloud protocol experiment and the open-ended survey questions will be thematically analyzed to uncover deeper insights. This data will reveal user concerns and experiences, such as reactions to security warnings and privacy permissions, together with participants' explanations for their preferences.
Recurring themes, like app permissions or web browser navigation, will be identified.

Quantitative data will be statistically analyzed to highlight significant user preferences and concerns. Comparisons will be made between concerns about malware in mobile apps versus web browsers, and preferences for offline access and ease of navigation.

Common themes, like the need for better security features in mobile apps or simpler web browsers, will be identified.

Examples of analyzed data include security warning frequencies, user comments on security measures like biometrics, statistical comparisons of malicious activity concerns, analysis of permissions granted or denied, concerns about data tracking, and usability observations like task completion times and navigation ease.

\section{Ethics}

Participants' consent will be obtained prior to participation. All personal data will be anonymized to ensure privacy and confidentiality. Prior to the commencement of the study, approval will be obtained from the Institutional Review Board. By combining experimental tasks with the think-aloud protocol and survey responses, this methodology aims to provide a comprehensive understanding of user preferences and concerns regarding mobile apps and web browsers.

\section{Expected Outcome}

This study aims to provide insights into users' preferences and concerns regarding mobile apps and web browsers for accessing online services.

We expect to identify security concerns, such as worries about malware in mobile apps and web attacks on web browsers. Privacy concerns are anticipated to center around permissions in mobile apps versus control over privacy settings in web browsers, such as "incognito" mode.
For example, people may prefer e-shopping through web browsers rather than mobile apps, because apps save payment methods by default, thus creating a threat of information theft in case the device is lost or stolen.

Usability preferences will likely highlight the functionality and offline access of mobile apps against the ease of navigation and lack of installation for web browsers. For example, people may prefer accessing services through a browser, to prevent overloading their storage, compared to mobile apps that consume storage~\cite{bijlani2021did}. 

\section{Anticipated Contribution}

The findings will inform developers on how to enhance security, privacy and usability of mobile apps and websites based on user concerns and preferences.

Additionally, the study will contribute to academic literature on user behavior, security, and privacy in both mobile and web contexts, providing valuable data for future research in usable security and privacy. This includes the potential to extend the research to other domains, such as comparing user perceptions of iOS apps versus the Safari browser.

\bibliographystyle{plain}
\bibliography{usenix2024_SOUPS}

@INPROCEEDINGS{8835286,
  author={Somé, Dolière Francis},
  booktitle={2019 IEEE Symposium on Security and Privacy (SP)}, 
  title={{EmPoWeb: Empowering Web Applications with Browser Extensions}}, 
  year={2019},
  volume={},
  number={},
  pages={227-245},
  doi={10.1109/SP.2019.00058}}

@INPROCEEDINGS{4725734,
  author={Taivalsaari, Antero and Mikkonen, Tommi and Ingalls, Dan and Palacz, Krzysztof},
  booktitle={2008 34th Euromicro Conference Software Engineering and Advanced Applications}, 
  title={Web Browser as an Application Platform}, 
  year={2008},
  volume={},
  number={},
  pages={293-302},
  doi={10.1109/SEAA.2008.17}}

@INPROCEEDINGS{8013420,
  author={Muehlstein, Jonathan and Zion, Yehonatan and Bahumi, Maor and Kirshenboim, Itay and Dubin, Ran and Dvir, Amit and Pele, Ofir},
  booktitle={2017 14th IEEE Annual Consumer Communications \& Networking Conference (CCNC)}, 
  title={Analyzing HTTPS encrypted traffic to identify user's operating system, browser and application}, 
  year={2017},
  volume={},
  number={},
  pages={1-6},
  doi={10.1109/CCNC.2017.8013420}}

@article{carroll2023human,
  title={Human-Browser Interaction: Investigating Whether the Current Browser Application’s Design Actually Make Sense for Its Users?},
  author={Carroll, Fiona},
  journal={International Journal of Human--Computer Interaction},
  pages={1--12},
  year={2023},
  publisher={Taylor \& Francis}
}

@inproceedings{10.1145/1125451.1125615,
author = {Ha, Vicki and Inkpen, Kori and Al Shaar, Farah and Hdeib, Lina},
title = {An examination of user perception and misconception of internet cookies},
year = {2006},
isbn = {1595932984},
publisher = {Association for Computing Machinery},
address = {New York, NY, USA},
url = {https://doi.org/10.1145/1125451.1125615},
doi = {10.1145/1125451.1125615},
booktitle = {CHI '06 Extended Abstracts on Human Factors in Computing Systems},
pages = {833–838},
numpages = {6},
location = {Montr\'{e}al, Qu\'{e}bec, Canada},
series = {CHI EA '06}
}

@article{du2018empirical,
  title={An Empirical Analysis of Hazardous Uses of Android Shared Storage},
  author={Du, Shaoyong and Zhu, Pengxiong and Hua, Jingyu and Qian, Zhiyun and Zhang, Zhao and Chen, Xiaoyu and Zhong, Sheng},
  journal={IEEE Transactions on Dependable and Secure Computing},
  volume={18},
  number={1},
  pages={340--355},
  year={2018},
  publisher={IEEE}
}

@inproceedings{zhang2019android,
  title={Android-based smartphone authentication system using biometric techniques: A review},
  author={Zhang, Xinman and He, Tingting and Xu, Xuebin},
  booktitle={2019 4th International Conference on Control, Robotics and Cybernetics (CRC)},
  pages={104--108},
  year={2019},
  organization={IEEE}
}

@inproceedings{sharevski2024exploring,
  title={Exploring Phishing Threats through QR Codes in Naturalistic Settings},
  author={Sharevski, Filipo and Mossano, Mattia and Veit, Maxime and Schiefer, Gunther and Volkamer, Melanie},
  booktitle={Symposium on Usable Security and Privacy (USEC) 2024},
  year={2024}
}

@article{song2021advanced,
  title={Advanced evasion attacks and mitigations on practical ML-based phishing website classifiers},
  author={Song, Fu and Lei, Yusi and Chen, Sen and Fan, Lingling and Liu, Yang},
  journal={International Journal of Intelligent Systems},
  volume={36},
  number={9},
  pages={5210--5240},
  year={2021},
  publisher={Wiley Online Library}
}

@article{berger2021crystal,
  title={Crystal ball: From innovative attacks to attack effectiveness classifier},
  author={Berger, Harel and Hajaj, Chen and Mariconti, Enrico and Dvir, Amit},
  journal={IEEE Access},
  volume={10},
  pages={1317--1333},
  year={2021},
  publisher={IEEE}
}

@inproceedings{xu2014evasion,
  title={An evasion and counter-evasion study in malicious websites detection},
  author={Xu, Li and Zhan, Zhenxin and Xu, Shouhuai and Ye, Keying},
  booktitle={2014 IEEE Conference on Communications and Network Security},
  pages={265--273},
  year={2014},
  organization={IEEE}
}

@article{berger2023breaking,
  title={Breaking the structure of MaMaDroid},
  author={Berger, Harel and Dvir, Amit and Mariconti, Enrico and Hajaj, Chen},
  journal={Expert Systems with Applications},
  volume={228},
  pages={120429},
  year={2023},
  publisher={Elsevier}
}

@article{gupta2017cross,
  title={Cross-Site Scripting (XSS) attacks and defense mechanisms: classification and state-of-the-art},
  author={Gupta, Shashank and Gupta, Brij Bhooshan},
  journal={International Journal of System Assurance Engineering and Management},
  volume={8},
  pages={512--530},
  year={2017},
  publisher={Springer}
}

@article{bijlani2021did,
  title={Where did my 256 GB go? A Measurement Analysis of Storage Consumption on Smart Mobile Devices},
  author={Bijlani, Ashish and Ramachandran, Umakishore and Campbell, Roy},
  journal={Proceedings of the ACM on Measurement and Analysis of Computing Systems},
  volume={5},
  number={2},
  pages={1--28},
  year={2021},
  publisher={ACM New York, NY, USA}
}

@article{berger2022problem,
  title={Problem-space evasion attacks in the Android OS: a survey},
  author={Berger, Harel and Hajaj, Chen and Dvir, Amit},
  journal={arXiv preprint arXiv:2205.14576},
  year={2022}
}

\end{document}